\documentclass[%
reprint,
 amsmath,amssymb,
 aip,
 jcp
]{revtex4-1}

\usepackage{graphicx}% Include figure files
\usepackage{dcolumn}% Align table columns on decimal point
\usepackage{bm}% bold math
\usepackage{color} %TO BE REMOVED LATER
\usepackage{soul} %TO BE REMOVED LATER
\usepackage{hyperref}% add hypertext capabilities
\begin{document}

%\preprint{APS/123-QED}

\title{Improving the ergodic characteristics of thermostats using higher order temperatures}% Force line breaks with \\

\author{Puneet Kumar Patra}
\affiliation{%
 Indian Institute of Technology Kharagpur, West Bengal, India 721302 
}%
\author{Baidurya Bhattacharya}%
 \email{baidurya@civil.iitkgp.ernet.in}
\affiliation{%
Indian Institute of Technology Kharagpur, West Bengal, India 721302 
}%

\date{\today}% It is always \today, today,

\begin{abstract}
Most deterministic schemes for controlling temperature by kinetic variables (NH thermsotat), configurational variables (BT thermostat) and all phase space variables (PB thermostat) are non-ergodic for systems with a few degrees of freedom. While for the NH thermostat ergodicity has been achieved by controlling the higher order moments of kinetic energy, the issues of nonergodicity of BT and PB thermostats still persist. In this paper, we propose a family of modifications for improving their ergodic characteristics. To do so, we introduce two new measures of configurational temperature (second and third order) based on the generalized temperature - curvature relationship. The equations of motion for the existing thermostats are modified by controlling the relevant higher order temperature variables. The ergodic characteristics of the proposed modifications are tested using a single harmonic oscillator. For the PB thermostat, the fastest route to ergodicity is by controlling the usual configurational and kinetic temperatures along with the $2^{\text{nd}}$ order configurational temperature. For the BT thermostat, controlling the usual configurational temperature along with the $2^{\text{nd}}$ order configurational temperature is sufficient to make it ergodic. Our method also provides a new ergodic generalization of NH thermostat that reduces to kinetic moments method for a single harmonic oscillator.
\end{abstract}

%Use showkeys class option if keyword
\maketitle

%\tableofcontents
\section{\label{Sec:Intro}Introduction}

Ergodicity of dynamics is a prerequisite for estimating statistical-mechanical properties from a single dynamical trajectory observed over a sufficiently long period of time \cite{ref56}. The ergodic hypothesis enables us to equate the time averages obtained from dynamical trajectories with the ensemble averages. Many real life experiments are performed under a constant temperature environment and thus, temperature control algorithms (or thermostats) are introduced in molecular dynamics simulations. Over the years several deterministic \cite{ref27,ref11,ref12,ref51,ref37,ref19,ref36,ref18,ref49,ref23, ref3, ref39, ref53} and stochastic \cite{ref1,ref16,ref35,ref45,ref31} thermsotats have been proposed.  Most of the deterministic thermsotats are a derivative of the extended system method,  first highlighted by Nos\'{e} \cite{ref37,ref38}. An extension of Nos\'{e}'s work by Hoover \cite{ref19} (also known as the Nos\'{e}-Hoover thermostat) is amongst the most commonly used thermostats. 

But, developing a thermostat that satisfies the second requirement i.e. the ergodicity of the dynamics has been a challenge. The thermostats usually \textit{assume} ergodicity in the extended system variables to show that the equations of motion sample from the canonical distribution \cite{ref7,ref5}. It has been shown that for systems with few degrees of freedom (e.g. single harmonic oscillator), this assumption is not valid \cite{ref41,ref21}. Two of the most popular ways to improve ergodicity in the NH thermsotat are by - (i) introducing two (or more) pseudo friction additive thermostat variables \cite{ref5, ref32, ref23}, and (ii) adding additional variables for controlling the fluctuations of the reservoir variables \cite{ref36}. However, very recently doubt has been cast on whether the latter method has a fully ergodic dynamics \cite{ref40,ref54}. 

The general expression of temperature in terms of an arbitrary scalar valued phase-space function, $B$, is \cite{ref42, ref6, ref30}: 
\begin{equation}
\dfrac{1}{k_B T} = \dfrac{\langle \nabla . \nabla B\rangle} {\langle \nabla H . \nabla B \rangle}
\label{eq:general_temp}.
\end{equation}
Substituting $B$ as the kinetic energy $\sum p_i^2/2m_i$ gives 
\begin{equation}
T_{\text{kinetic},1} = \dfrac{2}{3Nk_B} \sum \limits _{i=1} ^{3N}\dfrac{p_i^2}{2m_i}
\label{eq:kin_temp},
\end{equation}
while substituting $B$ as the potential energy, $\phi$, gives
\begin{equation}
T_{\text{config},1} = \dfrac{1}{k_B} \dfrac{\langle \vert \vert \nabla_{r_i} \phi \vert \vert^2 \rangle}{\langle \nabla_{r_i}^2 \phi \rangle}
\label{eq:config_temp}.
\end{equation}
For reasons that will be clear later, we use the suffix 1 in (\ref{eq:kin_temp}) and (\ref{eq:config_temp}). Most of the temperature control algorithms in MD simulations involve constraining just the kinetic part of the temperature $T_{\text{kinetic},1}$, (\ref{eq:kin_temp}). But, kinetic temperature based thermostats fail in several situations \cite{ref6a,ref10,ref13}, including flowing non-equilibrium systems, and as a result, thermostats based on controlling the configurational temperature, (\ref{eq:config_temp}), were proposed \cite{ref33,ref9, ref46, ref46, ref47, ref43}. Probably the most popular configurational thermostat is the Braga-Travis (BT) thermostat \cite{ref53}. 

\begin{figure*}
\includegraphics[scale=0.45]{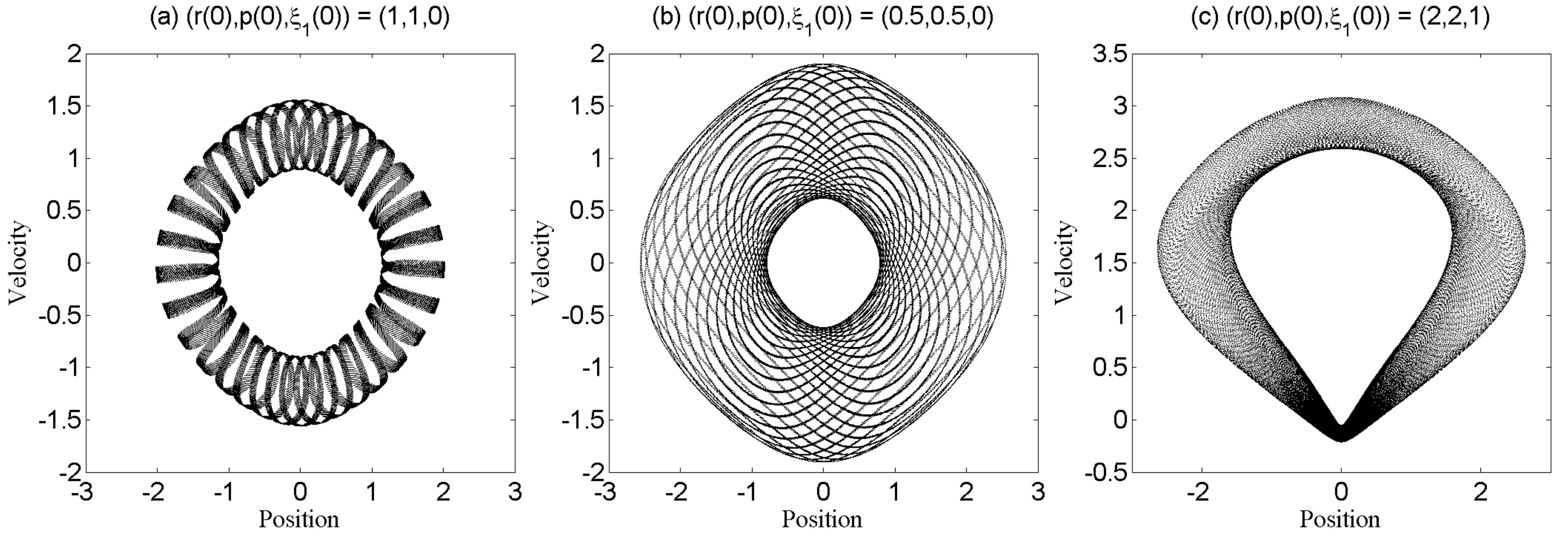}
\caption{\label{fig:BT_non_ergodicity} Non-Ergodicity of the BT thermostat. The phase-space plots (position-velocity) of the dynamics projected on to $\xi_1 = 0$ plane for three different initial conditions mentioned in the respective figures. The dynamics is not phase space filling. The presence of invariant tori, whose nature is dependent on the initial conditions, is self evident. Metric indecomposibility of phase-space, a necessary criteria for ergodicity to hold true, is not satisfied. The BT equations are solved using the fourth order Runge-Kutta method for 10 billion time steps, with $\Delta t = 0.001$.  }
\end{figure*}

For an equilibrium system, controlling either of the two temperatures is sufficient, since $T_{\text{kinetic},1} = T_{\text{config},1}$. Importantly, the equality also remains valid for a limited class of non-equilibrium systems that exhibit local thermodynamic equilibrium \cite{ref52, ref17, ref15}. The usual $T_{\text{kinetic}}$ and $T_{\text{config}}$ based thermosats are unable to ensure this equality \cite{ref24,ref26}. Hence, to cirumvent this problem, a new method of temperature control (PB thermostat) has been proposed recently \cite{ref39} that ensures the equality by simultaneous control of the kinetic and configurational temperatures. The PB dynamics is represented by the equations  
\begin{equation}
\begin{array}{ccl}
\dot{r_i} & = & \dfrac{p_i}{m_i} - \xi_{1} \dfrac{\partial \phi}{\partial r_i} ,\\ 

\dot{p_i} & = &\dfrac{\partial \phi}{\partial r_i} - \eta_{1} p_i,\\

\dot{\xi}_{1} & = &\dfrac{1}{Q_{\xi_{1}}} \sum \limits_{i=1}^{3N} \left[ \left( \dfrac{\partial \phi}{\partial r_i}^2 \right) - k_B T \dfrac{\partial^2 \phi}{\partial r_i^2} \right],\\

\dot{\eta}_{1} & = & \dfrac{1}{Q_{\eta_{1}}} \sum \limits_{i=1}^{3N} \left[ \left( \dfrac{p_i^2}{m_i} \right) - k_B T \right] ,
\label{eq:PB_thermostat}
\end{array}
\end{equation}
For a single harmonic oscillator with unit spring constant and mass (including $Q_{\xi_1}$ and $Q_{\eta_1}$), and kept at $k_BT = 1$, (\ref{eq:PB_thermostat}) has the form
\begin{equation}
\begin{array}{cc}
\dot{r} = p - \xi_{1}r, & \dot{p} = -r - \eta_{1} p, \\
\dot{\xi}_{1} = r^2 - 1, & \dot{\eta}_{1} = p^2 - 1.
\label{eq:SHO_PB}
\end{array}
\end{equation}
One can recover the equations of motion for the BT configurational thermostat by substituting $\eta_{1} = \dot{\eta}_{1} = 0$ \cite{ref53} and the NH thermostat \cite{ref19} by substituting $\xi_{1} = \dot{\xi}_{1} = 0$ in equation (\ref{eq:PB_thermostat}) and (\ref{eq:SHO_PB}). Unlike for NH thermostat, very few comprehensive attempts \cite{ref43} that deal with the ergodicity (or its lack thereof) of configurational thermostats have been made. Nevertheless, the BT equations are nonergodic, as shown in figure (\ref{fig:BT_non_ergodicity}).

\begin{figure*}
\includegraphics[scale=0.45]{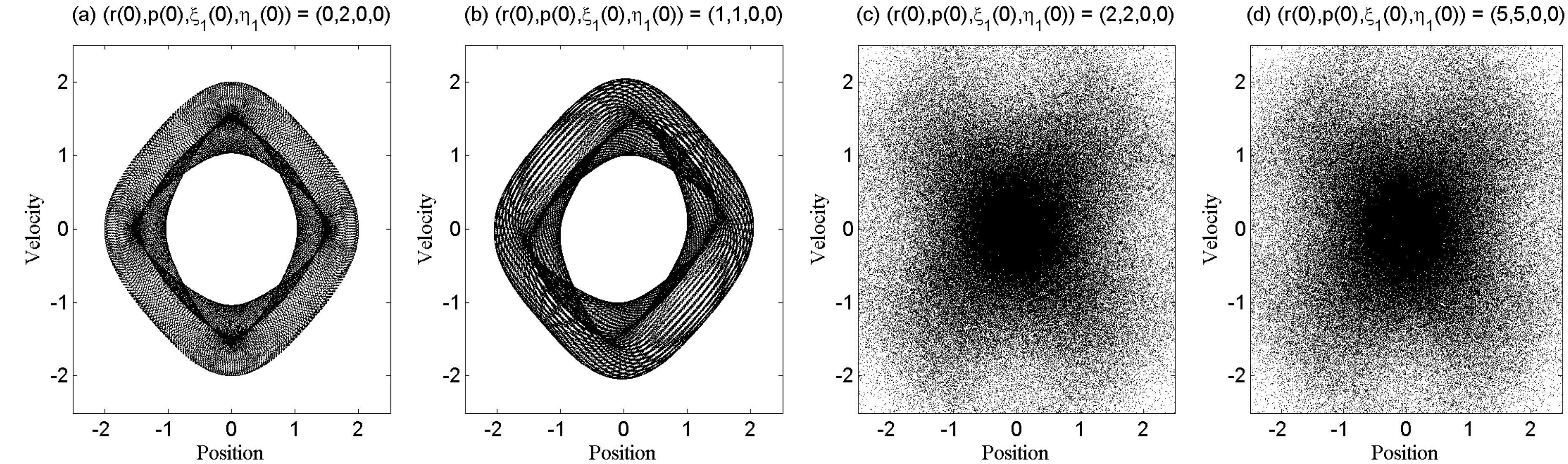}
\caption{\label{fig:pb_non_erg} Nonergodicity of the PB thermostat as evidenced from the projected phase space plots with four different initial conditions. In cases (a) and (b), invariant tori can be seen, while in (c) and (d), an $X$ shaped structure can be seen. The PB equations are solved using the fourth order Runge-Kutta method for 10 billion time steps, with $\Delta t = 0.001$. In cases (c) and (d), one may naively interpret that the dynamics is ergodic due to its phase-space filling nature. But, a look at the marginal distributions of position and velocity, indicates that the necessary (but not sufficient) condition for the dynamics to be ergodic - the marginal distributions must be Gaussian, is not satisfied.}
\end{figure*}

Likewise, the PB thermostat, despite being two-parametric \cite{ref48} and satisfying the general relationship of temperature control \cite{ref32}, is nonergodic \cite{ref44}, as shown in figure (\ref{fig:pb_non_erg}). Surprisingly, until very recently, it was thought that the fastest route to ergodic dynamics,  ensuring correct canonical sampling, is through two-parameter based thermostats \cite{ref48}. 

While ergodicity issues associated with NH thermostat are generally thought to be resolved (using the kinetic-moments method), they still persist for BT and PB thermostats. In this paper, our objective is to introduce modifications in the equations of motion of the BT and PB thermostats such that their ergodic characteristics improve. We do so by introducing two new measures of configurational temperature ($T_{\text{config},2}$ and $T_{\text{config},3}$), akin to the fourth and sixth moment relationship of kinetic temperature and velocity ($T_{\text{kinetic},2}$ and $T_{\text{kinetic},3}$, respectively). We develop a family of thermostatting equations that can simultaneously control $T_{\text{kinetic},1}$ through $T_{\text{kinetic},3}$ along with $T_{\text{config},1}$ through $T_{\text{config},3}$. The modified equations of motion are then subjected to a single harmonic oscillator. Results indicate that the modified equations drastically improve the ergodic characteristics of the originial thermostats. Our method also results in a new moments based generalization of the NH dynamics that reduces to kinetic-moments based thermostat for a single harmonic oscillator.

The paper is organized as follows: the new measures of configurational temperatures are introduced in section \ref{Sec:New_measures}, equations of motion for controlling the temperature variables are derived in section \ref{Sec:thermostat_development} and then the numerical simulations are shown.

\section{\label{Sec:New_measures} New Measures of Configurational Temperature}
We introduce two new measures of configurational temperature in this section that serve in modifying the BT and PB equations of motion. The usual kinetic temperature, $T_{\text{kinetic},1}$, is the scaled standard deviation of the velocity probability distribution. Due to the Gaussian nature of the velocity distribution, the kinetic temperature can be related to the higher order moments of velocity as well:
\begin{equation}
 k_B T_{\text{kinetic},2}= \sqrt{\dfrac{\langle p_i^4 \rangle}{3}}, \  
 k_B T_{\text{kinetic},3} = \sqrt[3]{\dfrac{\langle p_i^6 \rangle}{15}} .
\label{eq:kin_higher_order}
\end{equation}
We refer to these as the second order and third order kinetic temperatures, and hence the subsrcipts 2 and 3. It turns out that  (\ref{eq:kin_higher_order}) can be obtained through appropriate selection of $B$ in equation (\ref{eq:general_temp}): 
\begin{equation}
\begin{array}{ccc}
B=\sum \limits _{i=1} ^{3N} p_i^4/4 & \implies & T = T_{\text{kinetic},2} \\
B=\sum \limits _{i=1} ^{3N} p_i^6/6 & \implies & T = T_{\text{kinetic},3} \\
\end{array}
\end{equation}
In a similar way, we utilize the generalized relationship (\ref{eq:general_temp}) to introduce the higher order measures of configurational temperature. Selecting $B=\phi^2$, we get the second order configurational temperature $T_{\text{config},2}$:
\begin{equation}
\dfrac{1}{k_B T_{\text{config},2}} = \dfrac{\langle \vert \vert \nabla_{r_i} \phi \vert \vert^2 + \phi \nabla_{r_i}^2 \phi \rangle}{\langle \phi \vert \vert \nabla_{r_i} \phi \vert \vert^2 \rangle}
\label{eq:config_temp_second_order},
\end{equation}
and, by selecting $B=\phi^3$, we get the third order configurational temperature, $T_{\text{config},3}$,
\begin{equation}
\dfrac{1}{k_B T_{\text{config},3}} = \dfrac{\langle 2 \phi \vert \vert \nabla_{r_i} \phi \vert \vert^2 + \phi^2 \nabla_{r_i}^2 \phi \rangle}{\langle \phi^2 \vert \vert \nabla_{r_i} \phi \vert \vert^2 \rangle}
\label{eq:config_temp_third_order}.
\end{equation}
When $\phi = (1/2)r^2$, the similarity between the same orders of kinetic and configurational temperatures becomes apparent. In the next section we will utilize these higher order configurational and kinetic temperatures to develop equations of motion for simultaneously thermostatting these temperatures.

\section{\label{Sec:thermostat_development}Controlling higher order kinetic and configurational temperatures in simulations}
A consequence of nonoergodicity of dynamics is that the probability distribution function, hence the moments of the relevant phase-function estimatated from the time history, deviate from the corresponding ensemble averages (or the \lq{}\lq{true values}\rq{}\rq{}). Thermostats based upon standalone higher order kinetic temperatures obtained from different moments of kinetic energy have been developed in the late 1980s \cite{ref29}, but they also suffer from the problems of nonergodicity \cite{ref23}. The first breakthrough in search for moments based ergodic thermostats came through the kinetic moments method (HH) of Hoover and Holian \cite{ref23} that simultaneously controls the temperatures corresponding to the first and the second moments of the kinetic energy ($K$). Recognizing that the kinetic energy is distributed according to $\chi^2$ distribution, the kinetic temperature can be expressed in terms of second moment of $K$ as
\begin{equation}
k_BT = \dfrac{\langle 4K^2 \rangle}{\langle 2K \left(N+2\right) \rangle}
\end{equation}
It is interesting to note that one can find the same expression of temperature by substituting $B = K^2$ in (\ref{eq:general_temp}). The dynamics due HH thermostat are governed by the equations 
\begin{equation}
\begin{array}{cc}
\dot{r_i} = p_i, & 

\dot{p_i} = -\dfrac{\partial \phi}{\partial r_i} - \eta_{1} p_i - \eta_{2} (K/K_0) p_i,\\

\dot{\eta}_{1} = \dfrac{1}{Q_{\eta_{1}}} \left( K - K_0 \right), &

\dot{\eta}_{2} = \dfrac{1}{Q_{\eta_{2}}} \left( NK^2 - (N+2)KK_0\right),

\label{eq:HH_thermostat}
\end{array}
\end{equation}
where, $K_0 = Nk_BT/2$. The HH control removes the possibility of error from the fourth moment along with the second moment and thus shows marked improvement in ergodicity. Taking cue from the HH thermostat, we augment the PB dynamics by simultaneously and selectively controlling upto the third order kinetic and configurational temperatures. One can use the same methodology for controlling even higher orders as well. 

Let the contribution of the first three orders of the configurational and the kinetic temperatures be embedded in the dynamics through the variables $(\xi_{1},\xi_{2},\xi_{3})$ and $(\eta_1,\eta_2,\eta_3)$. The coupling between the system variables $(r_i,p_i)$ and the thermostat variables is sought to be of the form:
\begin{equation}
\begin{array}{ccl}
\dot{r_i} & = & p_i - \xi_1 \dfrac{\partial \phi}{\partial r_i} - 2 \xi_2 \phi \dfrac{\partial \phi}{\partial r_i} - 4 \xi_3 \phi^2 \dfrac{\partial \phi}{\partial r_i}, \\

\dot{p_i} & = & -\dfrac{\partial \phi}{\partial r_i} - \eta_1 p_i - \eta_2 p_i^3 - \eta_3 p_i^5, \\

\dot{\xi}_1 &=& ?,\ \ \ \   \dot{\xi}_2 = ?,\ \ \ \   \dot{\xi}_3 = ?, \\

\dot{\eta}_1 & = & ?,\ \ \ \   \dot{\eta}_2 = ?,\ \ \ \  \dot{\eta}_3 = ?.

\end{array}
\label{eq:Generalized_coupling}
\end{equation}

Our objective is to find the time evolution of the thermostat variables such that the extended phase-space distribution becomes canonical in all the variables, in the same manner as \cite{ref21,ref53,ref39}, i.e. the extended phase space density is: $f \propto \exp(-\beta H - 0.5c_{\xi_1}\beta \xi_1^2 - 0.5c_{\xi_2}\beta \xi_2^2 - 0.5c_{\xi_3}\beta \xi_3^2 - 0.5c_{\eta_1}\beta \eta_1^2 - 0.5c_{\eta_2}\beta \eta_2^2 - 0.5c_{\eta_3}\beta \eta_3^2)$, where $c_i$s are constants. To do so, the steady-state extended phase-space Liouville's equation is then solved (assuming statistical independence of the variables),
\begin{align}
& \dfrac{\partial f}{\partial t} + \sum\limits_i \left( \dot{r_i}\dfrac{\partial f}{\partial r_i} + \dot{p_i}\dfrac{\partial f}{\partial p_i} \right) + \sum\limits_j \left( \dot{\xi_j}\dfrac{\partial f}{\partial \xi_j} + \dot{\eta_j}\dfrac{\partial f}{\partial \eta_j} \right) \nonumber \\ 
& + f \left( \sum\limits_i \left( \dfrac{\partial \dot{r}_i}{\partial r_i} + \dfrac{\partial \dot{p}_i}{\partial p_i} \right) + \sum\limits_j \left( \dfrac{\partial \dot{\xi}_j}{\partial \xi_j} + \dfrac{\partial \dot{\eta}_j}{\partial \eta_j} \right) \right) = 0.
\label{extended_Liouville_Eqn}
\end{align}
The governing equations therefore become: 

%\begin{equation}
%\begin{array}{ccl}
%
%\dot{\xi_c} & = & \dfrac{1}{Q_{\xi_c}} \sum \limits_{i=1}^{3N} \left[ \left( \dfrac{\partial \phi}{\partial r_i} \right)^2 - \dfrac{1}{\beta} \left( \dfrac{\partial^2 \phi}{\partial r_i^2} \right) \right], \\ \\
%
%\dot{\eta_c} & = & \dfrac{1}{Q_{\eta_c}} \sum \limits_{i=1}^{3N} \left[ \phi \left( \dfrac{\partial \phi}{\partial r_i} \right)^2 - \dfrac{1}{\beta} \left( \phi \dfrac{\partial^2 \phi}{\partial r_i^2}  + \left( \dfrac{\partial \phi}{\partial r_i}\right)^2 \right) \right], \\ \\
%
%\dot{\psi_c} & = & \dfrac{1}{Q_{\psi_c}} \sum \limits_{i=1}^{3N} \left[ \phi^2 \left( \dfrac{\partial \phi}{\partial r_i} \right)^2 - \dfrac{1}{\beta} \left( \phi^2 \dfrac{\partial^2 \phi}{\partial r_i^2}  + 2\phi \left( \dfrac{\partial \phi}{\partial r_i}\right)^2 \right) \right], \\ \\
%
%\dot{\xi_k} & = & \dfrac{1}{Q_{\xi_k}} \left[ \sum _{i=1}^{3N} p_i^2 -\dfrac{3N}{\beta} \right], \\ \\
%
%\dot{\eta_k} & = & \dfrac{1}{Q_{\eta_k}} \left[ \sum _{i=1}^{3N} p_i^4 -\dfrac{9N}{\beta} \sum _{i=1}^{3N}p_i^2 \right], \\ \\
%
%\dot{\psi_k} & = & \dfrac{1}{Q_{\psi_k}} \left[ \sum _{i=1}^{3N} p_i^6 -\dfrac{15N}{\beta} \sum _{i=1}^{3N}p_i^4 \right].
%
%\end{array}
%\label{eq:thermostat_evolution}
%\end{equation}

\begin{equation}
\begin{array}{ccl}
\dot{r_i} & = & p_i - \xi_1 \dfrac{\partial \phi}{\partial r_i} - 2 \xi_2 \phi \dfrac{\partial \phi}{\partial r_i} - 4 \xi_3 \phi^2 \dfrac{\partial \phi}{\partial r_i}, \\

\dot{p_i} & = & -\dfrac{\partial \phi}{\partial r_i} - \eta_1 p_i - \eta_2 p_i^3 - \eta_3 p_i^5, \\

\dot{\xi}_1 & = & \dfrac{1}{Q_{\xi_1}} \sum \limits_{i=1}^{3N} \left[ \left( \dfrac{\partial \phi}{\partial r_i} \right)^2 - \dfrac{1}{\beta} \left( \dfrac{\partial^2 \phi}{\partial r_i^2} \right) \right], \\

\dot{\xi}_2 & = & \dfrac{1}{Q_{\xi_2}} \sum \limits_{i=1}^{3N} \left[ \phi \left( \dfrac{\partial \phi}{\partial r_i} \right)^2 - \dfrac{1}{\beta} \left( \phi \dfrac{\partial^2 \phi}{\partial r_i^2}  + \left( \dfrac{\partial \phi}{\partial r_i}\right)^2 \right) \right], \\

\dot{\xi}_3 & = & \dfrac{1}{Q_{\xi_3}} \sum \limits_{i=1}^{3N} \left[ \phi^2 \left( \dfrac{\partial \phi}{\partial r_i} \right)^2 - \dfrac{1}{\beta} \left( \phi^2 \dfrac{\partial^2 \phi}{\partial r_i^2}  + 2\phi \left( \dfrac{\partial \phi}{\partial r_i}\right)^2 \right) \right], \\

\dot{\eta_1} & = & \dfrac{1}{Q_{\eta_1}} \left[ \sum _{i=1}^{3N} p_i^2 -\dfrac{3N}{\beta} \right], \\

\dot{\eta_2} & = & \dfrac{1}{Q_{\eta_2}} \left[ \sum _{i=1}^{3N} p_i^4 -\dfrac{3}{\beta} \sum _{i=1}^{3N}p_i^2 \right], \\

\dot{\eta_3} & = & \dfrac{1}{Q_{\eta_3}} \left[ \sum _{i=1}^{3N} p_i^6 -\dfrac{5}{\beta} \sum _{i=1}^{3N}p_i^4 \right].

\end{array}
\label{eq:full_equation}
\end{equation}

The variables $Q_{\xi_i}$ and $Q_{\eta_i}$ can be viewed as mass of the $\xi_i^\text{th}$ and $\eta_i^\text{th}$ reservoir variable. It is easy to check that these equations of motion constrain (\ref{eq:kin_temp}), (\ref{eq:config_temp}), (\ref{eq:kin_higher_order}), (\ref{eq:config_temp_second_order}) and (\ref{eq:config_temp_third_order}). For a single harmonic oscillator of unit mass, potential $\phi = 1/2r^2$, unit thermostat mass and $\beta = 1$, (\ref{eq:full_equation}) can be written as:
\begin{equation}
\begin{split}
\dot{r} = p - \xi_1 r - \xi_2 r^3 - \xi_3 r^5, \\ 
\dot{p} = -r - \eta_1 p - \eta_2 p^3 - \eta_3 p^5, \\
\dot{\xi}_1 = r^2 -1, \dot{\xi}_2 = r^4-3r^2, \dot{\xi}_3 = r^6-5r^4 \\
\dot{\eta}_1 = p^2 -1, \dot{\eta}_2 = p^4-3p^2, \dot{\eta}_3 = p^6-5p^4 \\
\end{split}
\label{sho_equations}
\end{equation}
One can obtain different thermostats from the generalized equations (\ref{eq:full_equation}). The augmented form of the BT thermostat can be derived by substituting $\eta_1 = \eta_2 = \eta_3 = 0$ along with its derivatives in (\ref{eq:full_equation}). The augmented form of the NH thermostat can be derived by substituting $\xi_1 = \xi_2 = \xi_3 = 0$. 

For simplicity, we use the naming convention $C_i$ for only the $i^{\text{th}}$ order configurational temperature control, $C_{i,j}$ for the simultaneous control of the $i^{\text{th}}$ and the $j^{\text{th}}$ order configurational temperatures, and $C_{1,2,3}$ for the simultaneous control of the first three orders of configurational temperature. Analogously, for the kinetic temperature, the naming convention is $K_i$, $K_{i,j}$ and $K_{1,2,3}$. When both the configurational and kinetic temperatures are simultaneous controlled the naming convention is - $C_iK_m$ for $i^{\text{th}}$ order configurational and $m^{\text{th}}$ order kinetic temperature based control, $C_{i,j}K_m$ for the $i^{\text{th}}$ and $j^{\text{th}}$ order configurational temperatures along with the $m^{\text{th}}$ order kinetic temperature based control, $C_iK_{m,n}$ for $i^{\text{th}}$ order configurational temperature along with the $m^{\text{th}}$ and $n^{\text{th}}$ order kinetic temperatures based control and $C_{1,2,3}K_{1,2,3}$ for all the first three orders of configurational and kinetic temperatures based control. Using this style, the NH, BT and PB thermostats can be represented by $K_1$, $C_1$, and $C_1K_1$, respectively. $K_{1,2}$ control is not the same as that of HH thermostat, except for the case of a single harmonic oscillator, where both the equations are same. 

\section{\label{Improved_Ergodicity} Improving the ergodic characteristcs of the NH, BT and PB thermostats}
Ergodicity of dynamics is intrinsically linked with the metric indecomposibility of the phase-space, consequently for space-filling ergodicity no holes must be present in the dynamics \cite{ref41,ref34}. To circumvent the difficulty of gauging the presence of holes embedded within a higher (four and above) dimensional space from the projected dynamics, we look for holes instead in the dynamics at \textit{different Poincare sections}. Thus, we first see if there are any holes in the projected dynamics and if none could be detected, we check for the presence of holes at different Poincare sections \cite{ref40}. Another method to check ergodicity involves studying the difference between the maximum and minimum values of the largest Lyapunov exponents \cite{ref55, ref54}. If the difference is large, then system is non-ergodic. Ergodicity of the dynamics can also be confirmed by looking at the joint probability density functions (JPDFs) of the position and velocity at any Poincare section. If the dynamics is ergodic, the deviation of JPDFs from jointly bivariate standard normal will be small. In this study, we have utilized the Poincare section method along with the joint-normality of JPDFs (by comparing the marginal and joint moments) to establish (non)ergodicity.

We first show that the augmented NH equations are the same as the HH thermostat equations for a single harmonic oscillator and hence, ergodic. Following which we show that the Braga-Travis equations of motion result in marked improvement in the ergodic characteristics over the original BT equations. Subsequently, we show that the augmented PB equations also result in improved ergodic characteristics over the original PB equations.

\subsection{Improved ergodic characteristics of the augmented NH thermostat}
We now control the different orders of kinetic temperatures simultaneously by setting $\xi_i=\dot{\xi}_i=0$ in (\ref{eq:full_equation}). For a single harmonic oscillator of unit mass kept at unit temperature, and with all thermostats masses as unity as well, the augmented equations of motion are therefore,  
\begin{equation}
\begin{split}
& \dot{r}=p , \dot{p} = -r - \eta_1 p -\eta_2 p^3 - \eta_3 p^5,  \\ 
& \dot{\eta}_1 = p^2 -1, \dot{\eta}_2 = p^4 -3p^2, \dot{\eta}_3 = p^6 - 5p^4.
\label{eq:nh_full_sho}
\end{split}
\end{equation}
These equations of motion are identical to those obtained for the HH thermostatted single harmonic oscillator, and we conclude that the $K_{1,2}$ control improves the ergodicity of the oscillator.  

\subsection{Improved ergodic characteristics of the augmented BT thermostat(s)}
\begin{figure*}
\includegraphics[scale=0.4]{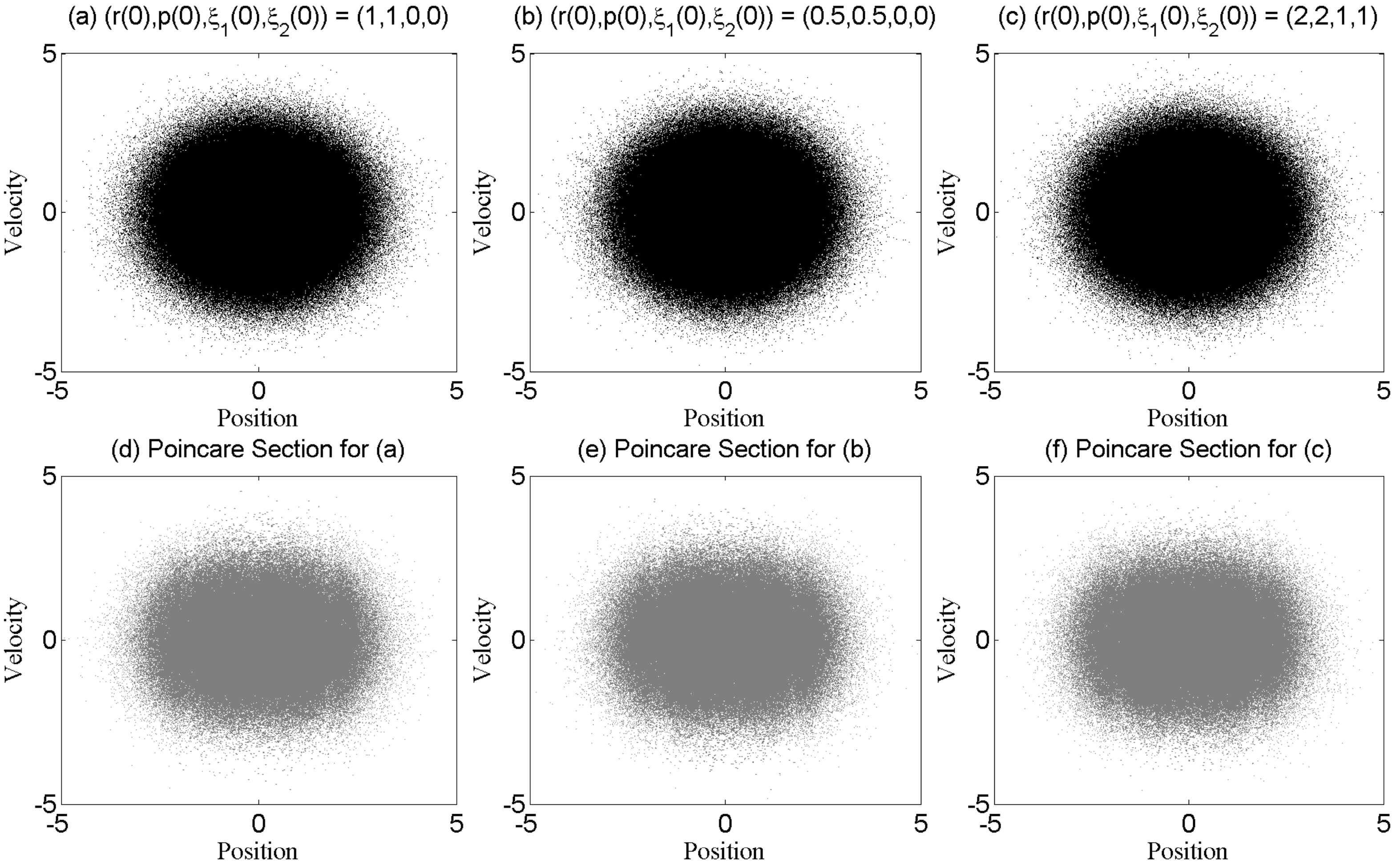}
\caption{\label{fig:two_four_control_bt} Improved ergodic characteristics of $C_{1,2}$ control. Top row (black) denotes the plot of the projected dynamics while the bottom row (gray) denotes the corresponding Poincare section plot  at $\xi_1 = 0$ and $\xi_2= 0$ plane. No existence of holes can be seen in any of the three cases. }
\end{figure*}

The different orders of configurational temperatures can be controlled simultaneously by setting $\eta_i=\dot{\eta}_i=0$ in (\ref{eq:full_equation}). For the single harmonic oscillator, the augmented equations of motion with $C_{1,2,3}$ control are therefore,  
\begin{equation}
\begin{split}
& \dot{r}=p-\xi_1 r - \xi_2 r^3 - \xi_3 r^5 , \dot{p} = -r,  \\ 
& \dot{\xi}_1 = r^2 -1, \dot{\xi}_2 = r^4 -3r^2, \dot{\xi}_3 = r^6 - 5r^4.
\label{eq:bt_full_sho}
\end{split}
\end{equation}
We begin with $C_{1,2}$ control. The equations of motion are solved using Runge-Kutta algorithm with $\Delta t = 0.001$ for 200 billion time steps. The projected phase space plots using three different initial conditions for this control along with the Poincare section at the $(\xi_1,\xi_2) = (0,0)$ plane are shown in Figure \ref{fig:two_four_control_bt}.  A comparison of Figures \ref{fig:BT_non_ergodicity} and \ref{fig:two_four_control_bt} suggests that the ergodic properties have improved marvellously by simply controlling an additional temperature variable. The dynamics, which previously was limited to a torus, now fills up the entire phase space. Additionally, there is no existence of any unoccupied space (hole) in the dynamics even at the Poincare sections.
To confirm if the dynamics is indeed ergodic, we also analyzed the joint and marginal densities of position and velocity for both the projected dynamics as well as the Poincare section. The results of the first three even marginal and joint moments of position and velocity indicate that their maximum deviation from the corresponding Gaussian distribution is less than 2.5\%. 

Thus, we see that due to additional control of the second order configurational temperature - (i) the entire phase space gets filled, and (ii) the distributions (marginal as well as joint) of position and velocity approach a Gaussian distribution. We, therefore, conclude that the ergodic characteristics of the augmented BT configurational thermostat, obtained by the simultaneous control of the first and the second order configurational temperatures, is much better than the original BT configurational thermostat . Similar arguments hold true when $C_{1,2,3}$ control is imposed.

\subsection{Improved ergodic characteristics of the augmented PB thermostat(s)}
There are 49 possible forms of augmented PB dynamics i.e. $C_iK_m$, $C_{i,j}K_{m}$, $C_{i}K_{m,n}$, $C_{i,j}K_{m,n}$, etc. To save computational efforts, we begin with checking if the ergodic characteristics show improvement with two-parametric thermostats, and then subsequently move to three-parametric thermostats.

\subsubsection{\label{PB_Non_ergodicity} Nonergodicity of $C_iK_m$ control}
\begin{figure*}
\includegraphics[scale=0.4]{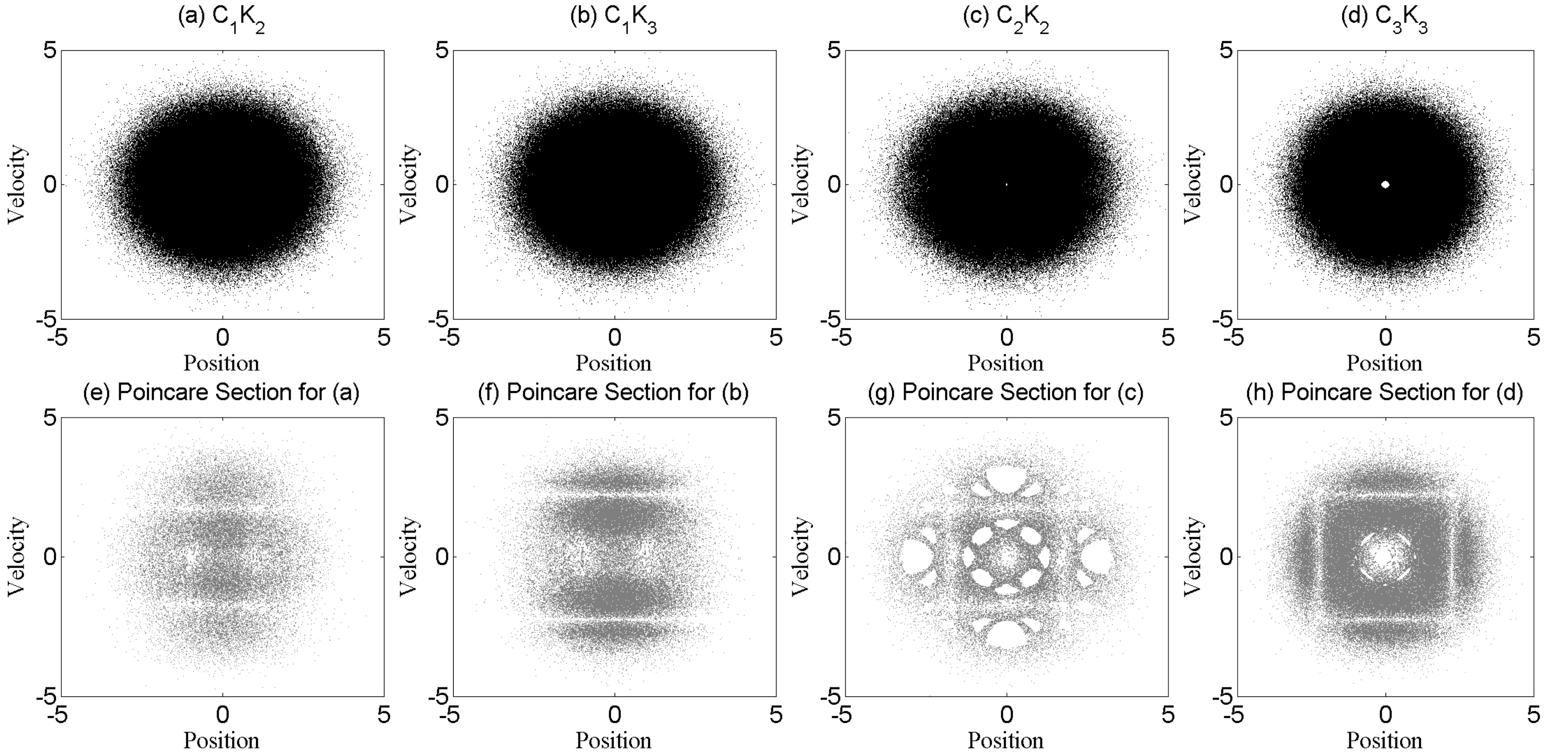}
\caption{\label{fig:pb_non_erg_ij_distributions} Nonergodicity of $C_iK_m$ control as evidenced from the phase space plots obtained using projected dynamics (top row) and at the Poincare section (bottom row) where thermostat variables are zero. Holes in the projected dynamics are present in (c) and (d), but cases (a) and (b) appear to be phase space filling. The Poincare section plot on the other hand shows that none of the four thermostatting conditions result in ergodic dynamics. The position and velocity are initialized at 1 for each case. All thermostat variables are initialized at 0 for (a) and (b), while for (c) and (d)  $\xi_i$ is initialized at 1.}
\end{figure*}

We have already seen the nonergodic behavior of the $C_1K_1$ control - the original PB thermostat. We find this to be true for any $C_iK_m$ control. Figure \ref{fig:pb_non_erg_ij_distributions} (top row) shows the resulting projected dynamics with $C_1K_2, C_1K_3, C_2K_2, C_3K_3$ controls. The bottom row indicates the position-velocity plot at the Poincare section where thermostat variables are zero. The equations are solved for 200 billion time steps with $\Delta t = 0.00005$ using classic $4^{\text{th}}$ order Runge-Kutta.
For $C_2K_2$ and $C_3K_3$ controls (Figures \ref{fig:pb_non_erg_ij_distributions} (c) and (d)), holes in the dynamics are present near origin in the projected dynamics, ruling out the possibility that the dynamics due to them is ergodic. Cases $C_1K_2$ and $C_1K_3$ are misleading if one looks just at the projected phase space plots (Figure \ref{fig:pb_non_erg_ij_distributions} (a) and (b)),  which indicate that the dynamics is phase-space filling. However, upon investigating their Poincare sections (Figure \ref{fig:pb_non_erg_ij_distributions} (e) and (f)), it is evident that the distribution of position-velocity in them are neither marginally Gaussian nor jointly Gaussian. Hence, the dynamics is not ergodic. 

\subsubsection{\label{PB_Ergo_Non_ergodicity} Improved ergodic characteristics of $C_iK_{m,n}$ and $C_{i,j}K_{m}$ controls}
\begin{figure*}
\includegraphics[scale=0.4]{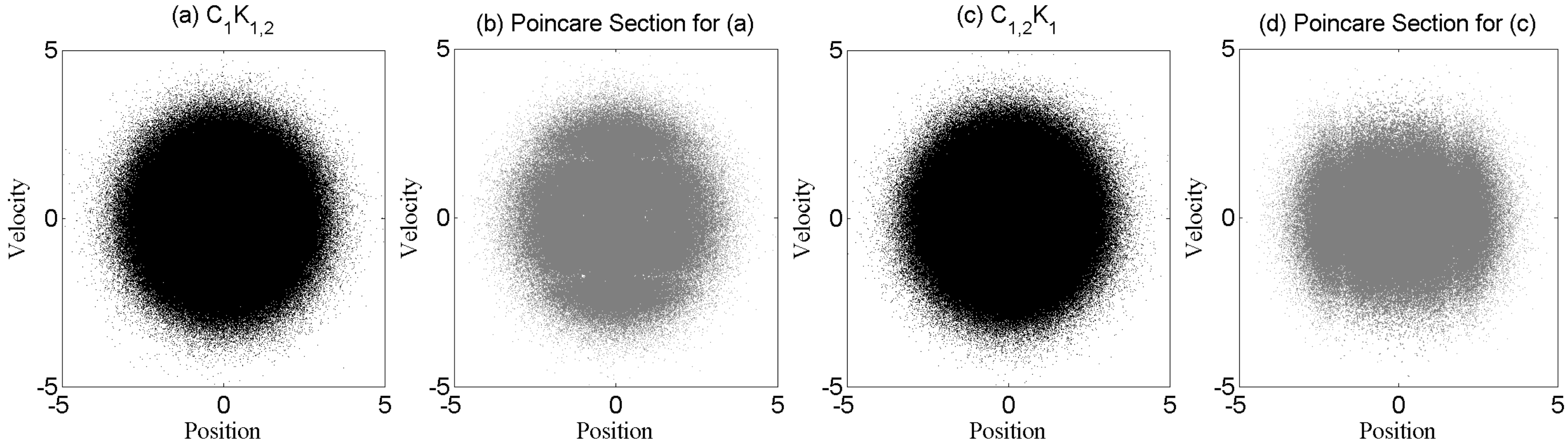}
\caption{\label{fig:pb_erg_ijk_distributions} Phase space plots obtained using $C_1K_{1,2}$ and $C_{1,2}K_1$ controls with $(r(0),p(0)) = (1,1)$ and all thermostat variables initialized at zero. The black figures, (a) and (c), indicate the position-velocity plot obtained using projected dynamics. The gray figures, (b) and (d), represent the Poincare section plot at the section where two thermostat variables are zero. The dynamics is phase-space filling.}
\end{figure*}

\begin{figure*}
\includegraphics[scale=0.4]{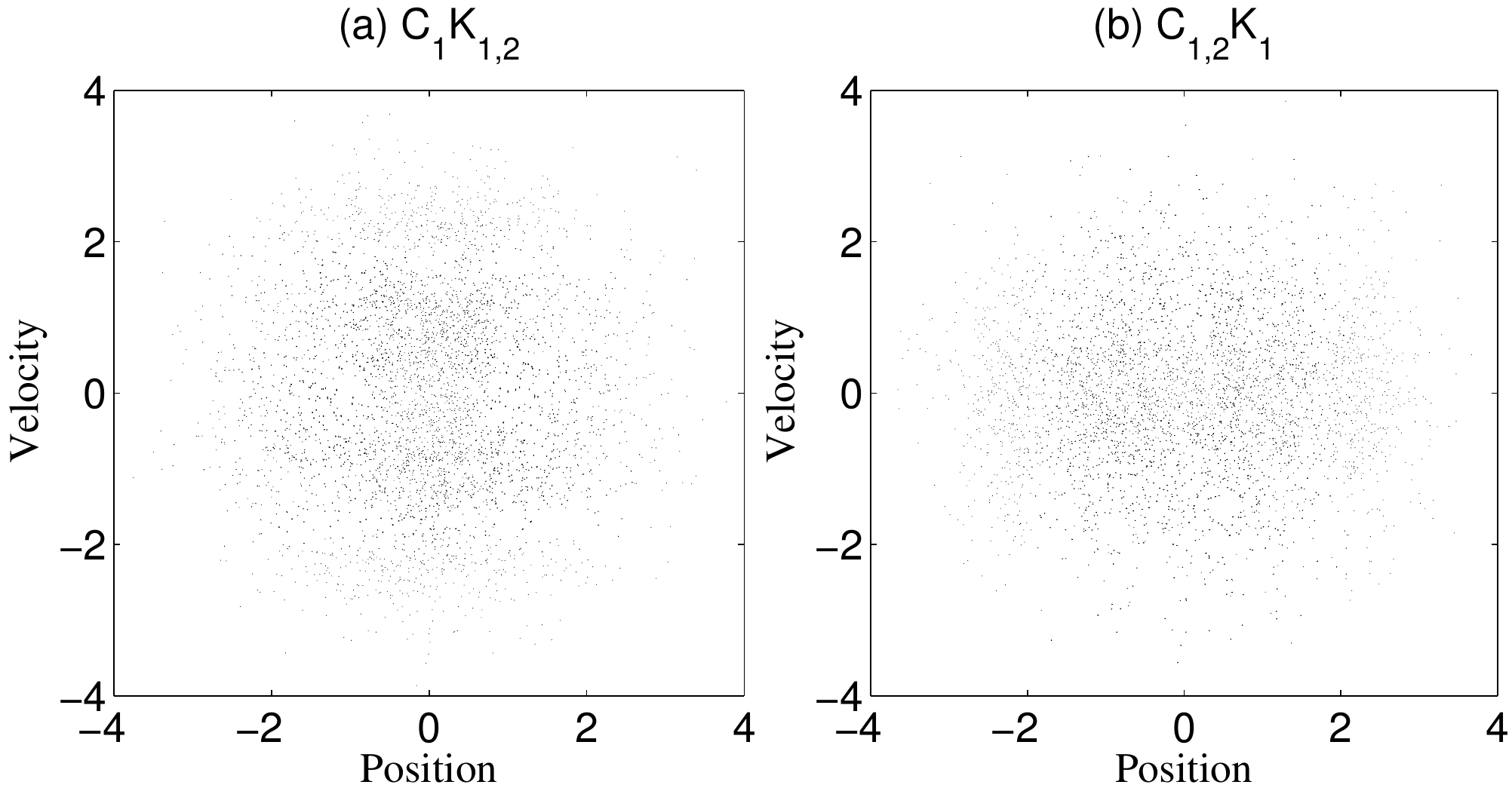}
\caption{\label{fig:pb_erg_ijk_distributions_ps} The triple Poincare section plots of velocity and position obtained using $C_1K_{1,2}$ and $C_{1,2}K_1$ with same initial conditions mentioned as in Figure \ref{fig:pb_erg_ijk_distributions}. The plots are obtained at the section where all three thermostat variables are zero.  Like before, there is no evidence of metric indecomposibility.}
\end{figure*}

Finally, we show that the $C_iK_{m,n}$ and $C_{i,j}K_m$ controls improve the ergodic characteristics of the PB dynamics. We begin with $C_1K_{1,2}$ and $C_{1,2}K_1$. The equations of motion are solved for 400 billion time steps, each of 0.0005. The projected phase-space plots along with the Poincare section plots (at section where two of the three thermostat variables are zero) for these controls are shown in Figure \ref{fig:pb_erg_ijk_distributions}, indicating that the dynamics is phase-space filling with no sign of any holes. For these controls, the same conclusion can be drawn from the position-velocity plots (Figure \ref{fig:pb_erg_ijk_distributions_ps}) of the triple Poincare section obtained at the section where all three thermostat variables are taken to be zero. Despite the small number of data present in it, there is again no sign of any holes in the phase space.


We use the data in Figure \ref{fig:pb_erg_ijk_distributions} to test if the resulting first three even order marginal and joint moments of position and velocity agree with those of a normal distribution. We do not use the data of the triple Poincare section for this purpose owing to the small number of data present in it (roughly 10,000 data points). The resulting marginal and joint moments (due to the projected dynamics and Poincar\'{e} sections) indicate that the maximum error for $C_1K_{1,2}$ method is 3.5\% while that of $C_{1,2}K_1$ control is 2\%. It is evident that the $C_{1,2}K_{1}$ control has a slightly better conformity to a normal distribution than the $C_{1}K_{1,2}$ control. Clearly, both the results are an improvement over the original PB formulation. A longer run is needed to check if the the results converge to a normal distribution. 

%\begin{table}
%\caption{\label{tab:table2} Comparison of the first three even order moments of the marginal and joint distributions of position and velocity for $C_1K_{1,2}$ and $C_{1,2}K_1$ controls. The projection column corresponds to the case where the distributions are obtained using the projected data. The PS column uses the distribution obtained from the data taken at the Poincare section where two of the three thermostat variables are zero. The deviation from a true normal distribution for botht the thermostatting cases are comparable.}
%\begin{ruledtabular}
%\begin{tabular}{|c|ll|ll|}
%\multicolumn{1}{l}{} & & & &  \\[\dimexpr-\normalbaselineskip-\arrayrulewidth]
%Moment & \multicolumn{2}{l|}{$C_1K_{1,2}$ control} & \multicolumn{2}{l|}{$C_{1,2}K_1$ control} \\ \hline
%
%& Projection & PS & Projection & PS \\ \hline
%
%${\langle r^2 \rangle}$    & 1.000 & 0.999 & 0.998 & 1.000 \\ \hline
%${\langle p^2 \rangle}$    & 1.000 & 1.000 & 1.003 & 1.001 \\ \hline
%${\langle r^2p^2 \rangle}$ & 1.001 & 0.998 & 0.998 & 1.002 \\ \hline
%${\langle r^4 \rangle}$    & 3.000 & 2.994 & 2.986 & 3.001 \\ \hline
%${\langle p^4 \rangle}$    & 2.996 & 2.991 & 3.018 & 3.009 \\ \hline
%${\langle r^4p^4 \rangle}$ & 9.044 & 8.884 & 8.951 & 9.099 \\ \hline
%${\langle r^6 \rangle}$    & 15.027 & 14.952 & 14.883 & 14.980 \\ \hline
%${\langle p^6 \rangle}$    & 14.983 & 14.899 & 15.172 & 15.092 \\ \hline
%${\langle r^6p^6 \rangle}$ & 231.485 & 217.160 & 226.708 & 229.501 \\
%\end{tabular}
%\end{ruledtabular}
%\end{table}
We similarly tested the ergodic characteristics of temperature control performed using four parameter based thermostat $C_{1,2}K_{1,2}$. Like the three thermostat control there is no evidence of any holes in the dynamics. We were unable to carry extensive moment based tests for this case owing to smaller number of data present. 

%\begin{figure*}
%\includegraphics[scale=0.4]{pb_2424_plots.eps}
%\caption{\label{fig:pb_2424_plots} The phase-space plots of position and velocity for the - (a) Projected dynamics, (b) Poincare section at $\xi_c=\xi_k=0$, and (c) Poincare section at $\eta_c = \eta_k = \xi_c = \xi_k = 0$ using $C_{1,2}K_{1,2}$ control with position and velocity initialized at 1 and thermostat variables initialized at 0. The equations are stiff and hence, solved for short time period using Matlab ode23s solver for only 1 billion time steps. In all the cases, there is no evidence of any holes in the dynamics.}
%\end{figure*}

\section{Conclusion}
The ergodic hypothesis enables us to equate time averages, obtained from molecular dynamics simulations, with ensemble averages. Lack of ergodicity in two of the thermostats - the Braga-Travis configurational thermostat (BT) and the recently proposed (PB) thermostat, limit their utility. In this paper, we introduce two new measures of configurational temperature ($T_{\text{config},2}$ and $T_{\text{config},3}$) through the generalized temperature-curvature relationship to augment the BT and the PB dynamics such that their ergodic characteristics improve. These higher orders of configurational temperatures are analogous to the kinetic temperature - velocity relations obtained using the fourth moment and the sixth moments of velocity. The equations of motion for the augmented PB dynamics are obtained by solving the steady-state Liouville's equation and are shown to satisfy the canonical distribution in the extended phase-space. A family of augmented PB thermostats have been obtained in which different orders of temperatures are controlled. The augmented form of BT thermostat can be obtained by dropping the terms needed for controlling the kinetic temperature. Alternatively, when the terms related to the configurational temperatures are dropped, we obtain the augmented form of the NH thermostat.

We subjected a single harmonic oscillator to the augmented NH, BT and PB thermostats. The augmented form of the NH thermostat for the single harmonic oscillator is identical to the HH thermostat, and therefore, possesses improved ergodicity. The augmented forms of BT thermostat are tested using $C_{1,2}$ and $C_{1,2,3}$ controls (not shown). The ergodic characteristics of these controls show marvellous improvement over the originial BT equations of motion. There are no holes in the phase space (both the projected as well as the Poincare section) and the deviation from canonical distribution is less than 2.5\%. 

For the augmented PB thermostat, simultaneous control of the configurational and kinetic temperatures can be done in many ways. Out of these, we found that none of the two-parameter based temperature control ($C_iK_m$)result in ergodic dynamics. The fastest route to ergodicity for augmented PB dynamics is through the $C_{1,2}K_{1}$ control. Ergodicity in the dynamics can be induced by using four parameter based thermostats as well. However, in some of the cases, the equations of motion become too stiff and require very small time step to be solved. It is possible that the best control would depend on the nature of the potential but we do not probe that angle further. 

%\section{Appendix}
%We show the phase-space plots of position-velocity using $C_1K_{1,3}$ and $C_1K_{2,3}$ temperature controls. The equations of motion corresponding to these temperature controls are too stiff. Consequently, one either needs either too small a time step for performing integration using standard Runge-Kutta techniques, or a stiff differential equation solver. We used MATLAB's inbuilt ode23s stiff differential solver for integrating the equations of motion for 100 million data points. The resulting phase space plots for the the Poincare sections $\eta_k=\psi_k=0$ and $\eta_c=\eta_k=\psi_k=0$ for $C_1K_{1,3}$ control and $\xi_k=\psi_k=0$ and $\eta_c=\xi_k=\psi_k=0$ for $C_1K_{2,3}$ controls are shown in Figure \ref{fig:pb_226_246_plots}. 
%
%\begin{figure*}
%\includegraphics[scale=0.4]{appendix_pb_226_246.eps}
%\caption{\label{fig:pb_226_246_plots} The phase-space plots of position and velocity for the $C_1K_{1,3}$ control (figures (a) and (b)) and $C_1K_{2,3}$ control (figures (c) and (d)). (a) is at the Poincare section $\eta_k=\psi_k=0$ while (b) is at the Poincare section $\eta_c=\eta_k=\psi_k=0$. (c) is at the Poincare section $\xi_k=\psi_k=0$ while (d) is at the Poincare section $\eta_c=\xi_k=\psi_k=0$. Again, we see no existence of any holes in the dynamics. However, no extensive moment tests could be performed due to small data available.}
%\end{figure*}

%\nocite{*}
\section{Acknowledgement}
The authors would like to thank Prof. William G. Hoover for his helpful comments.

\bibliography{apssamp}% Produces the bibliography via BibTeX.
\end{document}